\documentclass[journal]{IEEEtran}

\pdfoutput=1

\usepackage[utf8]{inputenc}
\usepackage[T1]{fontenc}

\usepackage[english]{babel}

\usepackage{lipsum}

\usepackage{color}

\usepackage[]{hyperref}
\hypersetup{
    colorlinks=true,
}

\usepackage{graphicx, subfigure} 
\usepackage[
    backend=biber,
    style=numeric
  ]{biblatex}
 
\begin{document}

\title{Towards Smart City Innovation\\ \large Under the Perspective of Software-Defined Networking, Artificial Intelligence and Big Data\\}


\author{Joberto S. B. Martins,~\IEEEmembership{IEEE Senior Member\\ Salvador University - UNIFACS}
\IEEEcompsocitemizethanks{
\IEEEcompsocthanksitem Prof. Dr. Martins, Joberto S. B. is with Salvador University - UNIFACS, Brazil. Invited Professor at HTW - Hochschule für  Techknik und Wirtschaft des Saarlandes, Saarbrucken - Germany. \protect\\
E-mail: joberto.martins@gmail.com}
\thanks{Manuscript received September 10, 2018; revised October 30, 2018. This work was supported by Salvador University - UNIFACS.}}

\maketitle

\markboth{RTIC - Revista de Tecnologia da Informação e Comunicação,~Vol.~8, No.~2,~Outubro~2018}{
Bhardwaj \MakeLowercase{\textit{et al.}}: Skeleton of IEEEtran.cls for Journals in VIM-Latex}



\begin{abstract}
Smart city projects address many of the current problems afflicting high populated areas and cities and, as such, are a target for government, institutions and private organizations that plan to explore its foreseen advantages. In technical terms, smart city projects present a complex set of requirements including a large number users with highly different and heterogeneous requirements. In this scenario, this paper proposes and analyses the impact and perspectives on adopting software-defined networking and artificial intelligence as innovative approaches for smart city project development and deployment. Big data is also considered as an inherent element of most smart city project that must be tackled. A framework  layered view is proposed with a discussion about software-defined networking and machine learning impacts on innovation followed by a use case that demonstrates the potential benefits of cognitive learning for smart cities. It is argued that the complexity of smart city projects do require new innovative approaches that potentially result in more efficient and intelligent systems. 

\end{abstract}

\begin{keywords}
Smart City, Innovation, Software-Defined Networking, Openflow, Artificial Intelligence, Machine Learning, Big Data, Cognitive Management, Resource Allocation.
\end{keywords}



\section{Introduction}
Cities present a clear trend to concentrate the majority of world's population in the years coming. The forecast is that about 65\% of world population will live in urban spaces by 2040 \cite{jalali_smart_2015}. City numbers are impressive: 30 cities will have more than 10 million inhabitants and largest cities will consume 85\% of world's energy and produce 80\% of the planet's waste \cite{bezerra_computational_2015}.

Easily perceivable problems in cities include congested traffic, unplanned urbanization, need for mobility, exploding population, scarce resources, waste management and energy distribution among others. 

A "Smart City" (SC), by definition, addresses the aforementioned urban problems looking for citizen quality of live improvement, promoting sustainability and engaging citizens through transparent government decisions \cite{martins_innovation_2017}.

There are various definitions for smart cities. Theolyre in \cite{theoleyre_networking_2015}, suggests that a smart city uses Information and Communication Technologies (ICT)  very intensively "to provide the ability to gather, analyze and distribute information so as to transform services, improve operational efficiency and entail better decisions." 

Smart city projects need to innovate \cite{martins_innovation_2017}. Innovation must be present due to the complexity, multidimensional characteristic, size and  multidisciplinary nature of the problems involved. Most of current legacy smart city solutions do not present the set of required capabilities or are not capable to cope with the ever increasing functionality complexity required by huge, efficiently managed and multipurpose urban spaces.

As such, innovation for the purpose of this discussion focused on city's smartness evolution can be understood as to evaluate and use a new set of emerging technologies that could effectively address in a efficient way the requirements and complexity of smart city's projects.

Government, institutions and private organizations need to plan their strategy towards a new smart city concept where innovation emerges as a fundamental requirement for both technical and managerial perspectives. It is expect that networks and computer technologies will keep playing a key role on this innovation process. 

More recently, a new set of network and computer paradigms and technologies have emerged. Software-defined networking (SDN), internet of things (IoT), network function virtualization (NFV), LTE/5G high speed wireless technologies, cloud computing, big data, cognitive management, machine learning and blockchain are examples of new elements that can bring a new perspective to smart city project developments.

\begin{figure}[htbp]
\begin{center}
  \includegraphics[scale=0.55]{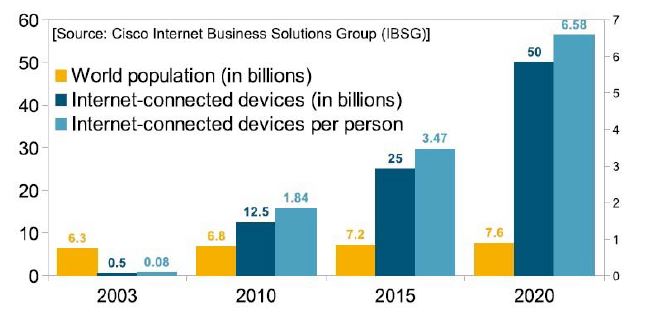}
\caption{Potentially Internet-connected devices and objects by user (Source: adapted from \cite{javed_internet_2018-1} and Cisco Internet Business Solutions Group)}
\label{fig:InternetGrowth}
\end{center}
\end{figure}

On the innovation path, another relevant aspect of smart city is the number of potentially internet-connected devices. In effect, most of the connected devices in smart cities are IoT sensors or actuators in some form (RFID tags, mobile phones, body sensors, ITS monitoring devices, others). In terms of numbers, it is expected to deal with millions or billions of sensors and actuators with multiple type of equipment by user (Fig. \ref{fig:InternetGrowth}).

This paper  aims to shed some light on how software-defined networking, artificial intelligence and big data may contribute either to create innovative solutions or to support developers and researchers on engendering new efficient solutions for smart city projects.

We identify initially the smart city project's main characteristics and issues. Following that we clarify the potential benefits introduced by these technological approaches in a scenario in which smart city projects are facing increasingly complexity and stringent requirements that are difficult to address only with either conventional or established technologies.

This paper is organized as follows. It initially discusses smart city projects requirements, challenges and proposes a simplified smart city deployment model to support the discussion. Following that, software-defined networking (SDN), artificial intelligence and big data are analyzed under the perspective of contributing to smart city innovation.

\section{Smart City Project Characteristics and Challenges}

A smart city project basically looks for improving the smartness of city's systems and applications and some of its requirements and characteristics include  \cite{bezerra_computational_2015} \cite{martins_innovation_2017} \cite{department_for_business_innovation__skils_smart_2013}:

\begin{itemize}
    \item A robust and scalable framework combined with secure and open access;
    \item A user-centric or citizen-oriented architectural approach;
    \item A huge volume of storable, findable, sharable, tagged, mobile and wearable public and private data enabling citizens to access information from anywhere and when needed; 
    \item An application level with analytically and integrated capabilities; and
    \item An smart physical and network infrastructure allowing the transfer of huge volume of heterogeneous data and the support of complex and distributed services and applications.

\end{itemize}

Smart city challenges are both technical and managerial. Managerial solutions are not discussed in this paper but do influence the technical perspectives discussed. For example, new business model and governmental strategies are required since smart city projects will bring, by definition, more effectiveness and transparency to most of current city legacy processes.

Another managerial aspect of smart city developments concerns the legacy "siloed" fashion in which projects are still developed by municipalities. In effect, the inherent complexity of smart city projects requires more integrated and systemic solutions. A multidisciplinary approach is most often the best one and this is a real challenge for most actual city administrations.

\section{Smart City Deployment Model}

\begin{figure*}[ht]
\centering
\includegraphics[width=1\textwidth]{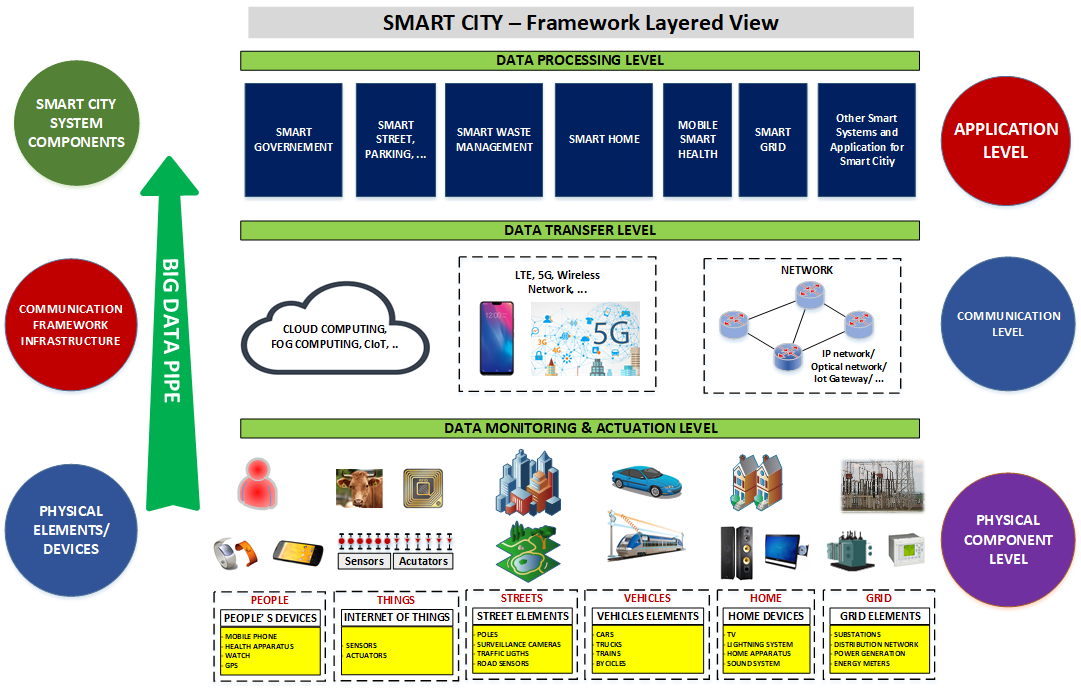}
\caption{Smart city deployment framework}
\label{fig:SCModel}
\end{figure*}

Smart city projects with a multidisciplinary perspective require a framework capable of structuring and modelling its basic components, its interactions and flow of data. The smart city's project deployment framework proposed for this discussion has a layered structure composed by three levels (Fig. \ref{fig:SCModel}):

\begin{itemize}
    \item A physical components level;
    \item A communications level; and
    \item An application level.
\end{itemize}

The physical components level hosts all devices, sensors, equipment, apparatus and components belonging to the various "smart" systems, services and applications.

The communications level comprises all network technological alternatives for providing communication facilities among physical elements and applications.

The application level includes all "smart" systems and developments conceived for the city. These systems are fed by and process a huge volume of data that uses the communication level to flow among the entities involved in the application process.

Each framework level presents a specific set of characteristics and requirements where innovation and new computer paradigms are being currently applied and researched.

At physical level the main issues are:
\begin{itemize}
    \item huge volume of devices;
    \item heterogeneity; and
    \item low-processing capabilities and energy limited resources.
\end{itemize}

At "application level" innovation also plays a significant role that is not addressed by this paper. In effect, it is at the application level where the integration provided by the physical and communication levels will produce the multidisciplinary effect that will impact the citizens quality of life.  Authors in \cite{santos_city_2018} \cite{hashem_role_2016} and \cite{jalali_smart_2015}  present and discuss examples on how new multidisciplinary and integrated smart city applications can benefit citizens.

The communication level is the main focus of this paper and its characteristics are highlighted and discussed in the next section.

\section{Network and Communication Infrastructure for Smart City with Innovation}

The communication framework infrastructure is a fundamental architectural element in any smart city project (Fig. \ref{fig:SCModel}). It has to address the urban context requirements with some unique characteristics and attributes like \cite{theoleyre_networking_2015} \cite{yaqoob_enabling_2017}:

\begin{itemize}
      \item High density of devices resulting mainly from the aforementioned trend on potentially internet-connected devices used by humans or automated systems and processes.
      \item A large and heterogeneous variety of networking technologies (HetNet) supporting sensors, actuators and devices that are present on the smart city deployment.
    \item A highly dynamic pattern of traffic to and from the physical elements, devices and applications. Traffic patterns generated by smart devices (IoT, sensors, other) can be either correlated or bursty. This is different of common traffic patterns generated by human beings with computers and, as such, is a new aspect to explore in terms of planning and dimensioning of resource allocation.
    \item A variety of physical topologies including unstructured and meshed ones.
    \item An interoperability inherent requirement since devices and applications must communicate.
\end{itemize}

In general the above mentioned characteristics and requirements make smart city projects "unique" requiring new approaches and, as such, is a potential field for innovation.

There is a variety of networking technologies, some innovative, that are currently being explored for use in smart city projects. Technologies like wireless sensor networks (WSN), machine to machine communication (M2M), vehicle to vehicle communication (V2V), cognitive radio  networks (CRN), visible light communication (VLC) and near field communication (NFC) are examples of innovative solutions being currently explored by researchers and developers \cite{yaqoob_enabling_2017}.

Having the smart city challenges, characteristics, basic requirements and deployment model in mind, the next question is how software-defined networking, artificial intelligence and big data  capabilities may contribute and foster smart city innovative projects.

\section{Why Software-Defined Networking for Smart City Deployment?}

\begin{figure}[htbp]
\begin{center}
  \includegraphics[scale=0.55]{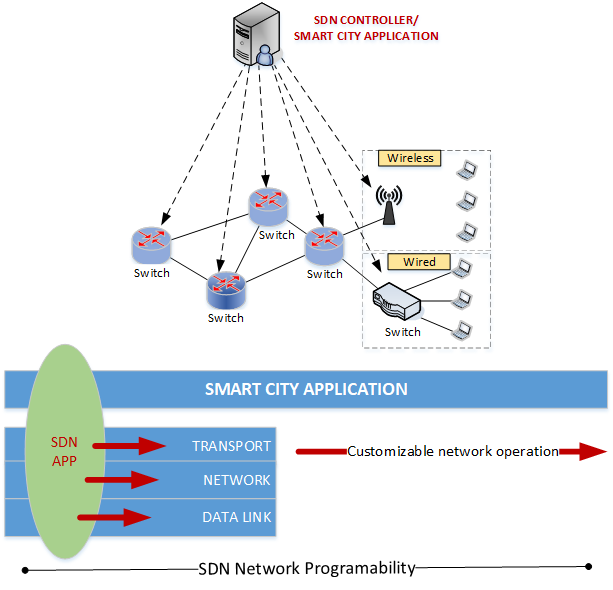}
\caption{SDN network programmability (Source: adapted from \cite{martins_innovation_2017})}
\label{fig:SDN}
\end{center}
\end{figure}

Software-defined networking (SDN) allows, in summary, the creation and deployment of programmable networks and systems \cite{kreutz_software-defined_2015}. In technical terms, SDN uses a logically centralized controller to program network equipment using a  well know interface  and protocol like Openflow (Fig. \ref{fig:SDN}). As a result, we can control network state transitions by monitoring network operational parameters and programming any modification of its operational behavior in terms of packet handling and manipulation \cite{martins_innovation_2017}.

SDN can fit and support smart city project developments in all three deployment levels:
\begin{itemize}
    \item at physical components level;
    \item at communications level; and
    \item at big data/ application level.
\end{itemize}

At communication level, smart city projects require the transfer of huge volume of data captured by heterogeneous sensors over large distributed areas. In this scenario, the communication infrastructure has to cope with heterogeneous communications requirements that are difficult to realize with a single or not dynamically configurable network infrastructure. Another important issue at the communication level is routing and traffic patterns variability. Routes have to be defined between data source and destinations and the network infrastructure has to adjust itself to the variability of communication resources demanded. Due to the variety of objectives and requirements involved in these communications such as QoS (Quality of Service), QoE (Quality of Experience) and SLA (Service Level Agreement) compliance, a flexible and programmable network infrastructure is the best possible approach.

SDN with Openflow, P4 or other deployment approach responds very positively to these difficult to achieve objectives and requirements. In effect, SDN adoption by smart city projects brings the following benefits:
\begin{itemize}
    \item Allows a logically centralized view and control of network equipment and resources; and
    \item Allows a dynamically configurable network infrastructure in terms of routing, where routing here means  the definition and setup of a path between data source and destinations. 
\end{itemize}

Data generated in smart cities will come from IoT devices, mobile phones, cars, e-health monitoring devices, video surveillance and smart grid equipment among many other possibilities. Each one of these specific "type"  of data will require distinct transfer requirements like QoS or any other similar performance parameter. As such, a configurable network infrastructure that can adapt itself to the various required requirements is a must for smart city innovative projects.

It is important to remark that SDN deployment in smart city projects do presents some important challenges. The most important one is the level of acceptance and maturity of the technology. Today's SDN deployment is prevalent in specific areas like private controlled networks, like Google's B4 approach for routing \cite{jain_b4_2013}, and the technology has not yet achieved a full consensus about its widespread utilization in all communication areas. Anyhow, our point is that network infrastructures have to evolve to a more dynamic and flexible operation and management approach and, for that, SDN is a suitable solution that brings various benefits for the deployed network. Beyond all this, SDN with virtualization has the capability and merit to keep legacy operation or, optionally, to virtualize its operation keeping both legacy an innovative networks operating simultaneously over the same equipment substrate.

For the communication level, SDN supports distinct solutions for the two basic networking approaches adopted in cities:
\begin{itemize}
    \item wired networking; and
    \item wireless networking.
\end{itemize}

Until very recently, most solutions based on SDN paradigm and its protocols (Openflow, P4, other)  have been designed for wired infrastructures \cite{ku_software-defined_2014}.

Nowadays, SDN-based solutions for wireless do prevail and have been extensively researched. This  comes from the fact that wireless in cities do support mobility and SDN allows a more accurate and centralized view and control of network states with mobile users.

As examples of wireless SDN-based solutions, Hakiri in \cite{hakiri_software_2017} explains why current protocols for mesh cloud do not fit new smart city requirements and SDN-based wireless solutions are presented. Rametta in \cite{rametta_s6_2017}, exploits SDN with the perspective of more easily deployment of services. This is achieved by merging SDN with network function virtualization (NFV) for a smart city video-surveillance system. The innovative results, according to authors is that "the video stream generated by each IP camera is automatically rerouted directly to the “interested receivers” in a point-to-multipoint fashion".

Another challenging research issue for the communication level (Fig. \ref{fig:SCModel}) in smart city projects is how to distribute the collected data among producers and consumers. Data will be mainly produced by IoT sensors and as such will be highly distributed, in large volume and with distinct requirements (QoS, QoE, SLA, other). Two technical aspects are involved in this discussion:
\begin{itemize}
    \item How semantically discover and distribute data; and
    \item How to allocate efficiently resources to the communication channels (communication level) necessary for data distribution.
\end{itemize}

One possible approach for the first challenge is currently being addressed by using Publish/ Subscribe approaches \cite{moraes_publish/subscribe_2018-2}. SDN plays again an innovative role with these approaches since it allows a more efficient control and deployment of network resources. Jalali in \cite{jalali_smart_2015} and Mazhar in \cite{mazhar_conceptualization_2015} present two distinct uses of SDN to foster resource allocation for IoT based smart city projects.

\section{How can Artificial Intelligence Contribute for Smart City Innovation?}

Artificial intelligence using machine learning techniques applied to smart city  has drawn a lot of attention from research. It has a tremendous application potential by exploring, for instance, a cognitive processing and management approach for networks, IoT and data.

The effective contribution that machine learning brings to smart city application and systems comes from the fact that:

\begin{itemize}
    \item A cognitive approach allows the computation of solutions for highly complex problems with multiple requirements and objectives.
    \item Cognitive solutions have the capability to extract "knowledge" and "learn" from huge and dynamic volume of data and this is an actual requirement to allow more "intelligent" and easy to use applications.
    \item Cognitive-based application and systems have the capability to substitute, at least in part, some human tasks (like management tasks), which turn  out to be difficult to realize and subject to errors when a large and sometimes unrelated volume of information is involved.
\end{itemize}

In relation to the smart city problem complexity, the computation of solutions under multiple requirements and objectives using an heuristic or meta-heuristic approach is difficult due to the NP-hard nature of the computation. Machine learning provides a new possible approach to solve the multiple requirement and objectives keeping computation on a reasonable level of difficulty and required computational capacity. This allows, for instance, the possibility to develop "on-the-fly" solutions for problems with a large number of decisions variables.

In smart city a large volume of data has to be captured, aggregated, smoothed, clustered and processed. Machine learning techniques do apply in this specific context mainly for clustering and knowledge extraction \cite{qiu_survey_2016} \cite{mohammadi_enabling_2018}.

In summary, artificial intelligence and machine learning are techniques known for decades that, nowadays, are capable to assist in the development of new solutions to many common problems found in smart cities for network infrastructure and IoT data. Machine learning allows intelligent network infrastructures with smarter data management and cognitive applications in general \cite{mohammadi_enabling_2018}.

\section{The Big Data Innovation Perspectives on Smart City Projects}

The big data innovation perspective for smart city projects under the technological point of view involves minimally the following aspects:

\begin{itemize}
    \item Big data analytics; and
    \item Big data processing, transfer and storage. 
\end{itemize}

First, it is important to highlight that the main target pursued by big data innovation in smart city is to extract information and knowledge efficiently to be used on behalf of the citizens.

Extract information and knowledge is certainly well accomplish by machine learning methods like reinforcement learning, case-based reasoning, deep learning, deep reinforcement learning (DRL) and unsupervised learning among other possibilities  \cite{gai_reinforcement_2018} \cite{mohammadi_enabling_2018}.

One challenging issue concerning big data within smart city projects is to decide "where" to process and store the big data chunk. The possibilities involved include:
\begin{itemize}
    \item To adopt an edge computing approach like used by fog computing \cite{mukherjee_survey_2018};  and/or
    \item To adopt either user-centric or content-centric approaches for data processing, communication and storage \cite{amadeo_information_2015} \cite{gai_reinforcement_2018}.
\end{itemize}

In all cases, introduce some "intelligence" is innovative in processing the big data flow from the physical components level to the application level through the communication level (Fig \ref{fig:SCModel}).

The technical aspects of big data processing, communication and storage are far extensive.  Innovation perspectives for the purpose of this specific discussion of big data issues in smart cities often converges to the adoption of edge computing or not for either user-centric or content-centric data processing approaches with big data analytics. In this specific context, both SDN and machine learning play a fundamental role.

Big data analytics is, by definition, well supported by machine learning with various alternative techniques (CBR, RL, deep learning, other) \cite{qiu_survey_2016}. 

Big data processing with SDN is discussed by  Khan concerning its communication and processing aspects in \cite{khan_big_2018}. Smart city big data processing and analytics by using different machine learning techniques is discussed by Mohammadi in \cite{mohammadi_enabling_2018} and Khan in \cite{hashem_role_2016}. Finally, edge computing has been extensively adopts in IoT scenarios with big data as discussed in \cite{mohammadi_enabling_2018}.


\section{Intelligent Communication Resource Allocation - A case Study}

A case study that highlights how innovation plays an important role in current cognitive management developments is presented in \cite{oliveira_cognitive_2018-1}.

The development aims to "enhance management" by adopting cognitive tools to support the overall management process (OAM - Operation, Administration and Management). The focus of this preliminary work is to control the allocation of bandwidth based on bandwidth allocation models (BAMs) operation for an MPLS network.

\begin{figure}[htbp]
\begin{center}
  \includegraphics[scale=0.6]{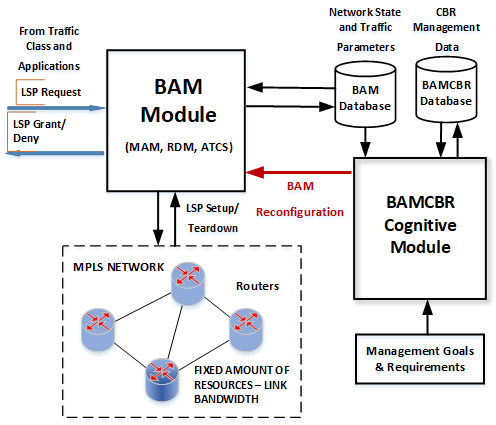}
\caption{Cognitive BAM-based bandwidth allocation (Source: adapted from \cite{oliveira_cognitive_2018-1})}
\label{fig:BAMCBR}
\end{center}
\end{figure}

The main components of the BAM-based cognitive approach are illustrated in Figure \ref{fig:BAMCBR}.

The overall operation, in summary, is as follows:

\begin{itemize}
    \item The BAM model controls bandwidth allocation per group of users in Traffic Classes (TCs);
    \item User request bandwidth to the BAM module that behaves as bandwidth broker granting or denying requests; and
    \item The BAM model used is dynamically configured as a function of input demands which are variable both in time and in terms of its bandwidth requirements (reflecting user's QoS/ QoE).
    
\end{itemize}

The result demonstrated by Reale in \cite{reale_preliminary_2014} and illustrated in Table \ref{tab:TrafficProfile} is that, depending on the input traffic pattern (profile of bandwidth demands) the BAM model must be reconfigured between distinct models (MAM, RDM or ATCS)  to optimize network operation like, for instance, to achieve the best possible link utilization. This reconfiguration is often mentioned as "BAM switching".

From the management point of view, BAM switching must be executed "on-the-fly" and the decision either to maintain actual configuration or reconfigure the BAM model has multiple requirements and possibly various objectives. Some possible management objectives include to increase link utilization, to reduce blocking or to reduce packet loss among others.

\begin{table}[ht]
\centering
\caption{Input Traffic Patterns (Source: adapted from \cite{oliveira_cognitive_2018-1}) }
\label{tab:TrafficProfile}
\begin{tabular}{|c|c|c|c|c|c|c|}
\hline
 \textbf{Traffic Profile} & \textbf{1}	& \textbf{2} & \textbf{3} & \textbf{4}	& \textbf{5} & \textbf{6}  \\ \hline
TC0 & High & Medium & Low & \multicolumn{3}{|c|}{High} \\\hline
TC1 & Low & Low & Medium & \multicolumn{3}{|c|}{High}\\\hline
TC2 & Low & High & High & \multicolumn{3}{|c|}{High}\\ \hline
Link Load & \multicolumn{3}{|c|}{$<$ 90\%} & \multicolumn{3}{|c|}{$>=$ 90\%} \\ \hline
Indicated BAM & RDM/ATCS & ALL & ALL & \multicolumn{3}{|c|}{MAM} \\ \hline
\end{tabular}
\end{table}

The main challenges involved in providing an efficient management solution for this bandwidth allocation setup are:

\begin{itemize}
    \item The MPLS network has a highly dynamic traffic pattern;
    \item MPLS users have, as often happens in actual networks, various quality of service (QoS) and quality of experience (QoE)  requirements; and
    \item The set of requirements for optimizing the configuration parameters of the BAM-based MPLS operation (bandwidth available per class, packet loss, delays, link utilization, other) is highly complex with multiple objectives.
    
\end{itemize}

The innovative approaches taken for this specific setup were:
\begin{itemize}
    \item To use case-based reasoning (CBR) and reinforcement learning (RL) machine learning techniques for BAM reconfiguration \cite{sutton_reinforcement_1998}; and
    \item To adopt SDN/Openflow for the deployment of the BAM-broker decisions in terms of LSP (Label Switched Path) setup, preemption or teardown over the MPLS network paths.
\end{itemize}

The option for using machine learning addresses the difficulty and challenge that we have to learn about network status and decide on reconfigure it or not. In effect, having multiple requirements and multiple objectives, as is the case, leads to a rather complex solution that are difficult to handle with heuristic or meta-heuristic based solutions. The machine learning approach presents effectively some difficulties in the modelling process but has the advantage of acquire knowledge and, as such, is capable to deal with much more complex scenarios reacting to them.

The adoption of SDN/Openflow is somehow disruptive, as the paradigm itself, and addresses the difficult we still have to dynamically configure networks. In effect, dynamic reconfiguration, like setting up and tearing down LSPs under dynamic requests is not evident on classical IP networks. An Openflow-based solution brings a number of valuable benefits like:

\begin{itemize}
    \item It allows a "global view" of the network state (established LSPs, link utilization, path setup, preemption and teardown, other) which is fundamental for any optimization process;
    \item It allows a "consistent" bandwidth allocation with network state changes under full control; and
    \item IT allows the maintenability of an external MPLS-like operation for users. In other words, the Openflow deployment emulates a classical MPLS behavior for external users.
\end{itemize}

Lessons learned from the adoption of machine learning and Openflow innovative techniques for the cognitive management of a BAM-based bandwidth allocation tool were \cite{oliveira_cognitive_2018-1}:

\begin{itemize}
    \item The machine learning cognitive management using CBR did learn from policies defined for the network and current performance metrics whether the current configuration is adequate and, subsequently, was able to dynamically and autonomically reconfigure BAM models to achieve the specified manager's goal.
    \item Network performance was improved in alignment with manager's predefined policy.
\end{itemize}

In summary, innovation was explored in two distinct ways:
\begin{itemize}
    \item Allowing the cognitive management of a system using a huge number of parameters that are difficult, if not impossible, to handle by humans (managers);
    \item Obtaining a "on-the-fly" solution that adapts the network configuration dynamically. By "dynamic" it is meant according with current user's traffic load over the network; and
    \item Deploying a new configuration "on-the-fly" which is not easily implementable by typical IP networks.
\end{itemize}

The first and second gains result from innovating with machine learning and the last one (network deployment) by using SDN/Openflow.


\section{Final Considerations}

Smart city projects present a complex new set of requirements with multiple objectives that involve multiple layers using heterogeneous network technologies for a huge number of users. Efficient solutions for smart city scenario are not easily achievable by current approaches typically based on heuristics and meta-heuristics.

Innovation based on new paradigm and technologies like SDN/Openflow and machine learning provides a new perspective to these projects by allowing the computation of solutions towards the cognitive management of complex network infrastructures coupled with the efficient processing of huge volume of data.





\printbibliography




%

\vskip -2\baselineskip plus -1fil

\begin{IEEEbiography}[{\includegraphics[width=1in,height=1.25in,clip,keepaspectratio]{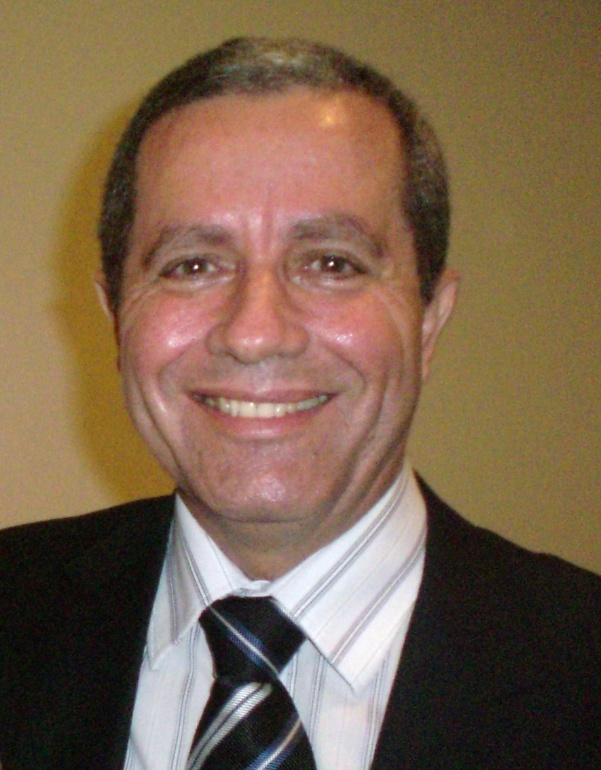}}]{Prof. Dr. Joberto S. B. Martins -}

Professor at Salvador University (UNIFACS) and PhD in Computer Science at Université Pierre et Marie Curie - UPMC, Paris (1986). Invited Professor at HTW - Hochschule für Techknik und Wirtschaft des Saarlandes (Germany) since 2003, Senior Research Period at Université of Paris-Saclay in 2016, Salvador University head and researcher at NUPERC and IPQoS  research groups on Resource Allocation Models, Software-Defined Networking - Openflow, Smart Cities, Smart Grid, Cognitive Management and AI application. Previously worked as Invited Professor at Université Paris VI and Institut National des Télécommunications (INT) in France and as key speaker, teacher and invited lecturer in various international congresses and companies in Brazil, US and Europe.

\end{IEEEbiography}





\end{document}